%% file: ms.tex
\documentclass[preprint2]{aastex}

\usepackage{epstopdf}
\usepackage{graphics, subfigure}
\usepackage{lscape}
\usepackage{psfig}

\input{defs.txt}

\newcommand{\nxs}{13}  
\newcommand{\ndla}{55}  

\begin{document}

\title{Unveiling the Secrets of Metallicity and Massive Star Formation  Using DLAs along Gamma-ray Bursts}

\author{
	A.~Cucchiara\altaffilmark{1},
	M.~Fumagalli\altaffilmark{2,3},
	M.~Rafelski\altaffilmark{1},
	D. Kocevski\altaffilmark{1},
	J.~X. Prochaska\altaffilmark{4},
	R.~J. Cooke\altaffilmark{4},
	G.~D. Becker\altaffilmark{5}
	}
	\email{antonino.cucchiara@nasa.gov}

\altaffiltext{1}{NASA Postdoctoral Program Fellow, Goddard Space Flight Center, Greenbelt, MD 20771, USA}
\altaffiltext{2}{Institute for Computational Cosmology, Department of Physics, Durham University, South Road, Durham, DH1 3LE, UK}
\altaffiltext{3}{Carnegie Observatories, 813 Santa Barbara Street, Pasadena, CA 91101, USA}
\altaffiltext{4}{Department of Astronomy and Astrophysics, UCO/Lick Observatory, University of California, 1156 High Street, Santa Cruz, CA 95064, USA}
\altaffiltext{5}{Kavli Institute for Cosmology and Institute of Astronomy, University of Cambridge, Madingley Road, Cambridge, CB3 0HA, UK}

\begin{abstract}
We present the largest, publicly available, sample of Damped Lyman-$\alpha$ systems (DLAs) along \swift\ discovered Gamma-ray Bursts (GRB) line of sights in order to investigate the environmental properties of long GRB hosts in the $z=1.8-6$ redshift range.
Compared with the most recent quasar DLAs sample (QSO-DLA), our analysis shows that GRB-DLAs probe a more 
metal enriched environment at $z\gtrsim3$, up to $[X/H]\sim-0.5$. In the $z=2-3$ redshift range, despite the large
number of lower limits, there are hints that the two populations may be more similar (only at 90\% significance level) than at higher redshifts.
Also, at \hiz, the GRB-DLA average metallicity seems to decline at a shallower rate than the QSO-DLAs:
GRB-DLA hosts may be polluted with metals at least as far as $\sim2\,$kpc from the GRB explosion site,
probably due to previous star-formation episodes and/or supernovae explosions. This shallow metallicity trend, extended now up to $z\sim5$, confirms previous results that GRB hosts are  star-forming and have, on average, higher metallicity than the general QSO-DLA population.
Finally, our host metallicity measurements are broadly consistent with the predictions derived from the hypothesis of two channels of GRB progenitors, one of which is mildly affected by
a metallicity bias, although more data are needed to constrain the models at $z\gtrsim 4$.

\end{abstract}


\section{Introduction }
\label{sec:intro}

One of the fundamental aspects of the formation of the first stars and galaxies is the actual 
conversion of the primordial hydrogen clouds into the first massive, almost metal free, objects 
\citep[Population III star,][]{Barkana:2001aa}. 
This first generation of stars, at $z\gtrsim 10$, disappeared quite rapidly due primarily
to strong negative feedback effects \citep[][]{Karlsson:2013aa,Bromm:2013aa,Yoshida:2008aa,Greif:2012aa,Chen:2014aa}.
However, some of these objects and the following generation of stars
probably ended their lives in very energetic explosions, either as pair-instability
supernovae or as long Gamma-ray Bursts \citep[GRB, see for a review][]{Meszaros:2013aa}, 
which can be detected by current and future high-energy missions up to the highest redshifts.

Thanks to the \swift\, satellite \citep{Gehrels:2004fj}, hundreds of GRBs have been discovered, even up to $z\approx8$ \citep{Tanvir:2009fv,Salvaterra:2009dz,Cucchiara:2011aa}. 
These \hiz\ GRBs can be used to test cosmic star-formation rate models as well as the cosmological chemical enrichment \citep{Kistler:2009aa,Robertson:2012aa,
Tanvir:2012aa,Salvaterra:2012aa,Ritter:2014aa}.
GRB progenitor models require massive, fast-rotating, and low-metallicity objects, \citep{Galama:1998aa,Woosley:1993aa,Hjorth:2003ab,Woosley:2011aa,Berger:2011aa,Levan:2014aa}, although the discovery of few GRBs that might have occurred in a  solar or even super solar metallicity environment challenges this paradigm \citep[e.g.][]{Prochaska:2009aa,Savaglio:2012ab,Kruhler:2012aa}.

Afterglow absorption spectroscopy of $z\gtrsim1.5$ GRBs provides a unique tool to determine the constituents of the GRB environment, in particular the amount of metals produced by past and on-going star-formation in the vicinity of the GRB explosion. This has important consequences for our understanding 
of the progenitor itself as well as the galaxies hosting GRBs throughout cosmic time.
GRBs are identified, at first, based only on their high energy emission, therefore their hosts can be studied in great detail after the afterglow emission disappears, representing a sample of star-forming galaxies unbiased with respect to their intrinsic luminosity. In fact, it is possible to use GRBs and their hosts to trace cosmic star-formation in an independent way compared  to magnitude limited Lyman-break galaxy surveys, although the effects of dust and metallicity biases are still under investigation \citep{Trenti:2014aa,Bouwens:2014aa,Levesque:2010aa,Modjaz:2008aa,Trenti:2013aa,Kocevski:2011aa}.

Similar to GRBs, quasars (QSO) have also been used for decades to study the effects of re-ionization, the conversion of neutral hydrogen into stars, and the cosmic metal enrichment. 
In fact QSOs, like GRB optical afterglows, are very bright and can be seen up to very high redshift.
The spectra of QSOs often show the presence of intervening absorbers (at redshifts lower than the QSOs), some of which are associated with large reservoirs of neutral hydrogen along their lines of sights. 
In particular, Damped Lyman-$\alpha$ systems (DLAs), by definition, have column density of neutral hydrogen 
 $\nh \geq 2 \NHunits$, while sub-DLAs  are defined as absorbers
with column density $10^{19} <\nh < 2 \NHunits$ (other types of subdivisions have been made, but they are not relevant for the purpose of this work).
These absorbers, which often trace galaxies along the QSO lines of sight \citep[e.g.][]{Fynbo:2011aa,Schulze:2012aa}, are the best laboratories to investigate the ISM, its evolution, and cosmic star-formation at high redshift, providing important constraints on galaxy 
evolution models \citep[e.g.][]{Wolfe:2005aa,Fynbo:2011aa,Neeleman:2013aa,Krogager:2013aa,Jorgenson:2014aa,Christensen:2014aa,Fumagalli:2014aa,Rafelski:2012aa}.

Recently, \cite{Rafelski:2014aa} have extended QSO-DLA studies up to $z\sim5$: they show that  the overall cosmological mean metallicity\footnote{The cosmic mean metallicity is defined as $\langle Z \rangle =\,$log\, $( \sum\limits_i 10^{[M/H]_i}N({\rm HI})_i)/ \sum\limits_i N({\rm HI})_i$, where $i$ is the redshift bin of DLAs as a function of redshift} slowly decreases from $z\approx1$ up to $z\approx 4.7$ \citep[see also][]{Jorgenson:2013aa,Prochaska:2003aa,Rafelski:2012aa} and then it appears to drop
rapidly below the extrapolated linear metallicity evolution, 
as if a sudden metallicity enrichment in DLAs occurs shortly after the end of re-ionization.
GRB-DLAs, in which the DLA is inside the GRB host, have been sparsely studied \citep{Fynbo:2008aa,Savaglio:2012ab,Sparre:2013aa,Arabsalmani:2014aa}, mainly due to the small sample size, the different data quality, and incompleteness. Nevertheless, \cite{Prochaska:2007ab} have derived a higher metal content at $z\gtrsim2$ for a set of GRB-DLAs with respect to a large sample of QSO-DLAs, suggesting that GRBs probe denser, more dust depleted, and metal rich regions then the QSO-DLAs population \citep[see also][]{Fynbo:2013aa}.

Finally, chemical evolution models suggest that DLAs metallicity measurements and relative abundance ratios at very \hiz\ enable us
to better understand the effect of the primordial PopIII stars chemical enrichment (and IMF) onto subsequent PopII stars IMF \citep{Salvadori:2012aa,Kulkarni:2013aa,Ritter:2014aa}.

The goal of this study is twofold: 1) compare the findings by \citet[][]{Rafelski:2014aa}, hereafter R14, with a large sample of GRB-DLAs that extends previous studies \citep[][]{Prochaska:2007ab,Savaglio:2012aa,Arabsalmani:2014aa}; 2) investigate the metallicity evolution of GRB-DLAs and compare it with host galaxy metallicity predictions at different redshifts.

The paper is structured as follows: in the next section we will present our samples of GRB- and QSO-DLAs; in
\S\ref{sec:analysis} we will describe our analysis, in \S\ref{sec:discussion} we will discuss our findings and 
possible biases, and in \S\ref{sec:conclusions} we will summarize our results. Throughout the paper we adopted the solar metallicity measurements from \cite{Asplund:2009aa}, with solar abundance of [X/H]$_{\odot}$ equal to 7.12 (Sulphur), 7.51 (Silicon), and 4.56 (Zinc).    


\section{Samples}
\subsection{Gamma-ray Burst DLAs}
We select our GRB-DLAs sample from all the GRB afterglows observed 
during the 2000-2014 time span for which \HI\ and metallicity measurements can be 
obtained (Table \ref{tab:list}). In order to detect the \Lya\, absorption line (1216\,\AA\ rest-frame) 
with most of the current spectrographs, a GRB has to be at least at $z\sim 1.8$ in order for the line to 
be redshifted out of the atmospheric blue cut-off (which 
usually means a minimum observed wavelength limit of $\sim$3400 \AA).
GRB-DLA absorbers are unambiguously associated with the GRB host galaxies, 
since often fine-structure transitions (e.g. \FeII* $\lambda2316$) or the termination of \Lya\ forest are identified at the same redshift of the \Lya\ feature \citep{Prochaska:2006aa,Vreeswijk:2007aa}.
The presented sample includes spectra obtained with different instrument resolutions: from low resolution
(resolving  power $R\sim400$) spectra obtained with the AlFOSC camera, to high resolution ($R\sim 55000$) obtained with the UVES instrument. 
Due to the transient nature of GRB afterglows, some spectra were obtained when the afterglow was quite faint and the resulting  signal-to-noise of the acquired spectra is not uniform within the sample. We therefore exclude GRBs with $S/N\leq3$ (at usually 6000\,\AA), due to the unreliable detection of metal lines (see Section \ref{sec:analysis}). 
All the data obtained with the Very Large Telescope (VLT)
instruments were retrieved from the ESO Archive\footnote{Based on data obtained from the ESO Science Archive Facility} or
were available within our collaboration\footnote{http://grbspecdb.ucolick.org/}. In few other cases we cannot obtain the raw data
and we include the results from abundance analysis as they appear in the literature (e.g. GRB\,050904 obtained with the FOCAS camera on the Subaru telescope).
We include these systems in Table \ref{tab:list} and briefly describe each line of sight in Appendix \ref{app:appA}.

Our sample includes \nxs\ GRB afterglow spectra obtained with the X-Shooter instrument mounted on the VLT. 
In order to analyze these data we primarily use our own customized pipeline written in {\tt IDL} (Becker et al., private comm.) as it is optimized for point sources and has an improved sky subtraction procedure. Additionally, we used the official pipeline (version 2.5, within the {\tt REFLEX} workflow \citep{Freudling:2013aa, Goldoni:2006aa} to verify the output of our custom pipeline. For completeness, in the last column of Table \ref{tab:list} we report the literature reference where these data, if published, appear.

Among the GRBs in Table \ref{tab:list} there are 12 sub-DLAs and we exclude them from the subsequent 
analysis due to the fact that these absorbers may probe a different environment than the general DLA population. These objects will be studied in a companion paper (Cucchiara et al. in prep), where we will present a more detailed description of the ionization field in order to reproduce the observed absorption pattern \citep[see for example][]{Vreeswijk:2013aa}.
We also exclude 4 GRBs where only upper limits on the \nh\, could be placed (GRB\,071020, GRB\,051111, and GRB\,080913, GRB\,090323), but we report them for completeness.
For other 4 GRB-DLAs only \NH\ measurement were obtained but the $S/N$ is too low for reliably identifying any metal feature (GRB\,020124, GRB\,060522, GRB\,080603B, and GRB\,121201A). 
In summary our GRB-DLA sample comprises \ndla\ GRB-DLA lines of sight (3 from the literature).
Among these \ndla,  we present metal abundances for 11 new GRB-DLA lines of sight.

\subsection{Quasar DLAs}
Thanks to the Sloan Digital Sky Survey large samples of quasars have been obtained up to $z
\sim7$. Thousands of these QSOs present DLAs along their lines of sight.  We use the current most complete list of high-resolution QSO-DLA spectra obtained by \citet[][hereafter R12]{Rafelski:2012aa} and R14, which extend previous work from, e. g., \citep{Prochaska:2003aa}. All these QSOs have been observed with high-resolution spectrographs and have very high S/N, which are of great importance in order to resolve multiple narrow metal feature (see also Section \ref{sec:resol}) as well as to provide accurate metallicity measurements using different metal tracers. 


\section{Analysis}
\label{sec:analysis}

Our dataset allows us to assess ionic metal abundances directly from the afterglow spectra rather than relying on the
literature. First, we select the spectra obtained 
with high-resolution spectrographs (which we defined at $R\gtrsim 6000$, see also \citealt{Jorgenson:2013aa}) and high S/N, and compare them to the QSO-DLA sample 
that has been obtained with high-resolution instruments only (e.g. HIRES on the Keck telescopes). 
Second, as already presented by \citet{Prochaska:2006ab}, abundance estimates (as well as the associate statistical 
error) obtained using low-resolution spectrographs 
are often under-estimated because of saturation effects, while blending may cause an over-estimation,
and therefore the measurements should be cautiously used. 
As mentioned before, we perform our own analysis for the GRB line of sights and, in the following sections, we discuss 
in detail our results from the low-resolution sample. Our approach yields a more homogeneous sample, similar to 
the QSO-DLAs sample, for which the same procedure has been applied.

\subsection{Low Resolution Sample}
\label{sec:resol}
\begin{figure*}[t!]
\epsscale{1.0}
\includegraphics[scale=0.70,angle=0]{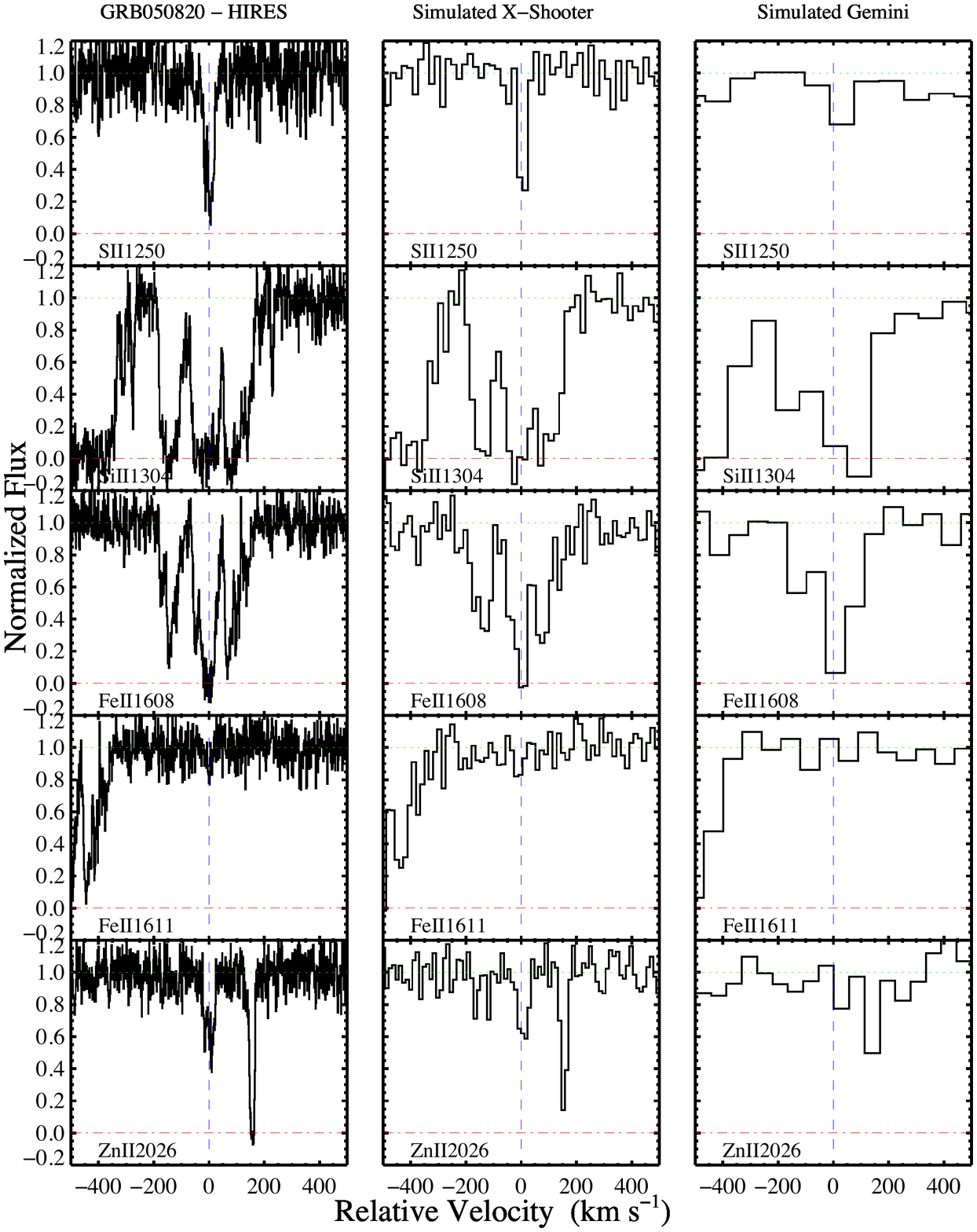}
\caption{\footnotesize{Left column shows weak metal lines identified in the Keck/HIRES ($R=30000$) spectrum of GRB 050820A, which has true metallicity of $Z/Z_{\odot}=+0.17$. Center and right columns show the same lines resampled at the typical X-Shooter ($R\sim6000$ in this case) and Gemini ($R\sim1200$) resolutions respectively, assuming a S/N=10. Green line represent the continuum level. In case of metal rich systems, like this example, reliable abundances can still be derived from the X-Shooter data, though blending can be an issue, while only limits can be placed from the Gemini data. Lower resolution instruments are even more heavily affected, making it almost impossible to determine meaningful limits on the metal abundances.}}
\label{fig:lincomp1}
\end{figure*}

\begin{figure*}[t!] 
\epsscale{1.0}
\includegraphics[scale=0.70,angle=0]{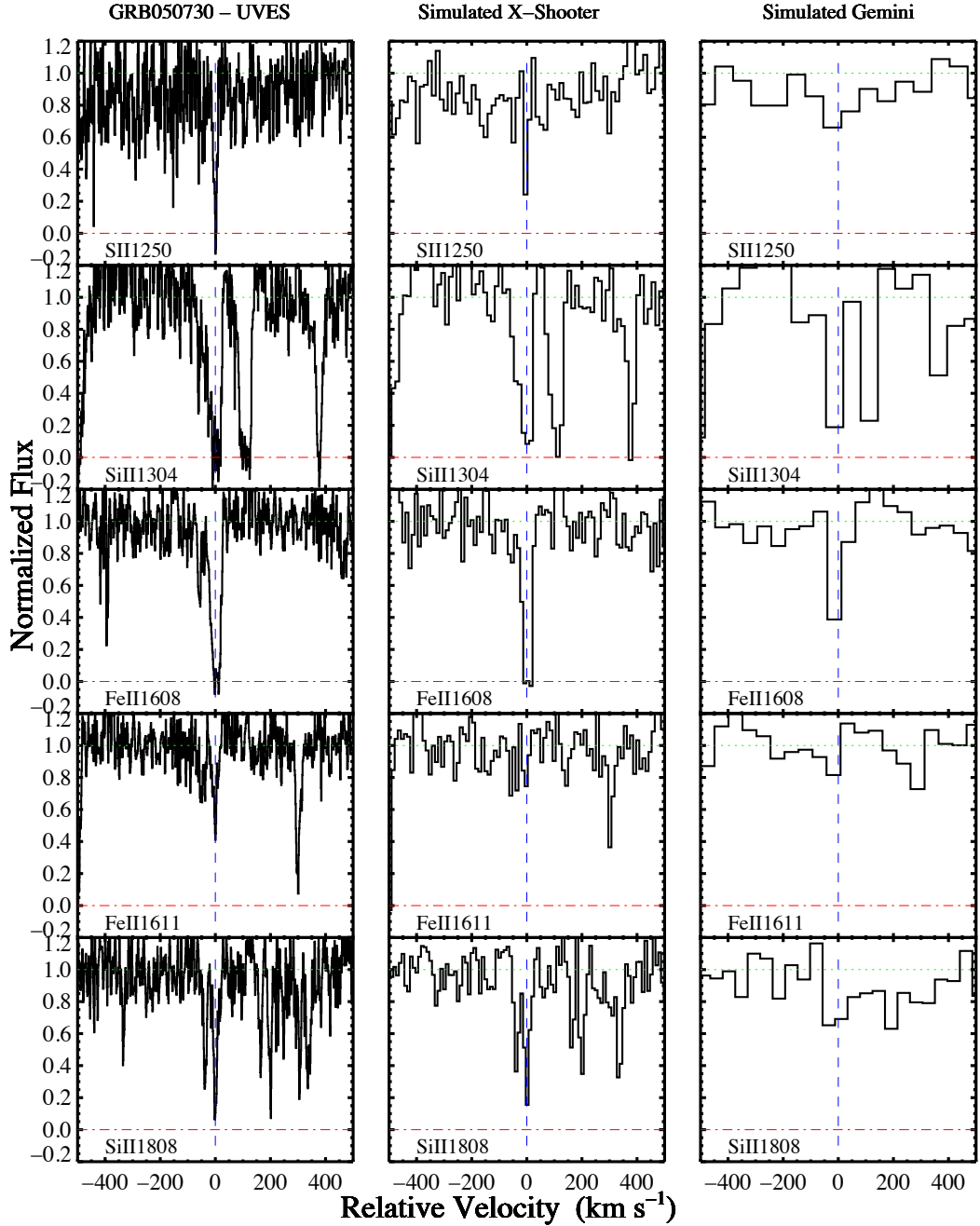}
\caption{\footnotesize{Same as Figure \ref{fig:lincomp1}, this time for one of the lowest metallicity systems GRB 050730 (obtained with the 
VLT/UVES spectrograph, $R\sim40000$) which has a true metallicity of $Z/Z_{\odot}=+0.01$. In this case it is obvious that the typical lines used for metal abundance measurements are also very narrow and almost disappear even at the X-Shooter resolution. This plot shows that measuring metallicity of the order of $Z/Z_{\odot}=+0.01$ or less is very difficult and often only limits can be placed (e.g. \SII1250 or \FeII1611).}}
\label{fig:lincomp2}
\end{figure*}

The consequences of low-resolution ($R\lesssim 6000$) spectroscopy on GRB-DLA abundance measurements via, e.g., Curve of Growth 
methodology \citep[COG,][]{Spitzer:1978aa} has been already discussed in the literature \citep{Prochaska:2006ab,Jorgenson:2013aa}. Our large and diverse sample enables us to directly compare
 abundance measurements obtained for systems with different intrinsic metallicity. In the first column of Figures \ref{fig:lincomp1} and  
 \ref{fig:lincomp2} we show line profiles for two systems observed with the VLT/UVES instrument which provides a resolution of 7 \kms\, (left column): GRB\,050820A (Figure \ref{fig:lincomp1})
 is a GRB-DLA with intrinsic moderate metallicity ($Z/Z_{\odot}=-0.76$), while GRB\,050730 is a metal poor system ($Z/Z_{\odot}=-1.96$)). In the second and third column we 
 present the same spectra resampled at the resolution of our average X-Shooter and Gemini/GMOS instruments respectively. 
Three important effects are evident and require particular attention when estimating abundances with mid/low-resolution instruments:
1) blending of nearby lines, either from other transitions rising from the GRB-DLA system or from unrelated intervening systems, may occur in lower resolution spectra (this also makes the determination of the continuum level difficult, and even more so with low S/N data) 2) hidden saturated lines in low-resolution spectra may not be clearly identified and thereby yield a lower value then the true abundance; 3) strong absorption features associated with moderately metal rich systems are still detected at $R\sim1200$ (or 200 \kms, GMOS typical resolution), but they completely disappear or are difficult to distinguish at even lower resolution, providing un-interesting metallicity abundance limits.

In the first case, the result is an overestimation of metal abundances and de-blending procedures may be very complexed, especially when trying to asses further hidden saturation  of the blended components. In the second, instead, the column density is underestimated. We therefore opted to use isolated weak lines (e.g. \FeII\,1608), or very strong
transitions (e.g. \SiII\,1526) which are likely saturated and therefore provide reasonable lower limits.
In the third case it is impossible to distinguish between a true low-metallicity system and the effect of 
low-resolution instrumentation. 

Since the majority of our spectra (34 over 55) 
have been obtained with $R\lesssim 2400$ and  the remaining spectra with $R\gtrsim 6000$ we adopt the latter as the minimum
resolution for which we can derive accurate metallicity estimates.
Therefore, for all the spectra with resolution lower than the X-Shooter spectrograph (typically $R\lesssim 6000$) and $S/N>3$, we measured metallicity from strong lines and provided only lower limits for the ionic abundances. 
We note that this analysis is similar to \cite{Jorgenson:2013aa}, where ionic abundances from an even lower resolution instrument, like the MagE spectrograph on the Magellan telescope ($R\approx 4200$), was compared with the X-shooter, UVES and HIRES instruments. 
Also, while ``hidden'' saturation can be still present, we carefully choose features with depths such to minimize this effect \citep{Prochaska:1996aa,Penprase:2010aa}, typically with normalized flux values $F_{\lambda} < 0.5$ in any pixel in the line profile.
\citet{Jorgenson:2013aa}  have also preformed several simulations on the reliability of using the Apparent Optical Depth method technique in deriving ionic abundances in such spectra  \citep[AOD,][]{Savage:1991aa} in comparison with Voigt profile fitting procedure.  
In order to further check this consistency we also performed a Voigt profile fitting for several ions in our X-Shooter and UVES spectra and compare these
values with the ones obtained by our AOD analysis. The column densities agree with each other on average within one standard deviation.
For these reasons (agreement between the AOD and Voigt profile methods, the fact that the majority of our spectra have $R\lesssim 6000$, and 
that our comparison sample of QSO-DLAs metal abundances are also obtained with the AOD technique), we present our metallicity measurements derived with the AOD methodology in Table~\ref{tab:list}.

\subsection{Neutral Hydrogen}
For all the GRBs in Table \ref{tab:list} for which we were able to retrieve the afterglow spectra, we 
determine the redshift of the GRB-DLA based on the simultaneous identification of the strong \Lya\ 
feature (identifiable also at low resolution) and at least one of the fine-structure transitions (like FeII* and NiII*) often present in GRB afterglow spectra \citep{Prochaska:2006aa,Vreeswijk:2007aa,Prochaska:2007ab}.
For those cases in which fine-structure lines  were not observed, e.g. due to spectral coverage, we required,
 that at least other low ions transitions were detected at the same redshift of \Lya, or that the end of the \Lya\ forest was also identified.

We fit the \Lya\ profile with a Voigt profile using the  {\tt x\_fitdla} procedure within the {\tt XIDL}\footnote{\tt http://www.ucolick.org/$\sim$xavier/IDL/} package, while we adopted the measurements derived in the literature if the spectra were not available (see notes in Table  \ref{tab:list}).

The QSO-DLAs measurements were obtained directly from R12 and R14, which performed a similar analysis on a sample of QSO spectra obtained with high-resolution spectrographs  on the Keck telescope (e.g. HIRES or ESI).

In Figure \ref{fig:hdistr} we present the column density distribution (f$_{\rm HI}$(N,z)) of our two samples in comparison with the model by \citet{Noterdaeme:2009aa} \cite[see also][]{Wolfe:1995aa,Peroux:2003aa,Prochaska:2009ab,Noterdaeme:2012aa}.
As previously noted \citep{Reichart:2002aa,Savaglio:2003aa,Jakobsson:2006aa,Prochaska:2007ab},  \HI\ column density in QSO-DLAs is a factor of ten lower than the GRB-DLAs \citep[although see][]{Noterdaeme:2012ab,Noterdaeme:2014aa} . This suggests that GRB-DLAs may trace a denser ISM phase, more similar to the sites of ongoing star formation \citep{Fynbo:2008aa}, while QSO-DLAs may probe, instead, a lower column density medium, possibly further from dense molecular regions.
Moreover, a handful of GRBs with large amount of neutral hydrogen ($\nh\ \gtrsim 10^{21.5} \cmsq$) along their line of sights exhibit the presence of molecular hydrogen (\H2) \citep{Prochaska:2009aa,Kruhler:2013aa,DElia:2014aa}, while only in a few cases have been found along quasars \citep{Noterdaeme:2008aa,Srianand:2012aa,Jorgenson:2013aa,Jorgenson:2014ab}.

\subsection{Metallicity}
Measuring the gas metal content in GRB-DLA systems is not simple, especially because some of the metals might be locked into dust grains \citep[dust depletion effect, ][]{Savaglio:2003aa}. For example, the mildly refractory element silicon is usually depleted in the Galaxy, but only marginally in DLAs and therefore it can often be used to determine the gas metallicity \citep{Wolfe:2005aa,Rafelski:2012aa}, assuming that this also applies for GRB-DLAs. 
Other good metallicity tracers include sulfur and zinc. 
Zinc, in particular, is often preferred because it is un-depleted in the ISM and has two strong transitions
at rest-frame 2026\,\AA\, and 2063\,\AA. However, it is only a trace element and therefore represents a small fraction of the 
mass density of the heavy element.  Moreover, the evolution of zinc resembles iron only for specific star formation histories and careful 
modeling of zinc production by SN II and SN Ia shows an underproduction of Zn and Mg compared to S, invoking additional production
site, such as intermediate-mass stars, in order to reconcile their abundance values\citep[see][for a detailed descritpion]{Fenner:2004aa}.
Therefore, as pointed out by \cite{Rafelski:2012aa} and \cite{Prochaska:2006ab}
we decided to use low-ionized transitions of sulfur (e.g. \SII\,1250), silicon (e.g. \SiII\,1808), and iron (e.g. \FeII\,1611), in order of importance. 
We only use the zinc lines in the few cases where the previously discussed lines are unavailable, such as when they fall at the location of atmospheric telluric bands.

We measured ionic abundances using the Apparent Optical Depth method \citep{Savage:1991aa}, which relies on the identification of several unsaturated lines of the same species and provides accurate measurements with no assumptions on the features Doppler parameter as other methods \citep[like Curve-of-Growth analyses,][]{Carroll:1996aa}. As we mention in the previous earlier these values well agree with the Voigt profile fitting technique.

In some cases, due to the presence of saturated transition, we only estimate lower limits on the overall metallicity \citep[see also][]{Savaglio:2004aa,Savaglio:2006aa,Prochaska:2006ab}. 
In fact, to be conservative we consider all our measurements from  low-resolution data as lower limits. 

In Table \ref{tab:list} we summarize our findings, including \HI\ and metallicity measurements, 
as $[X/H]$ relative to solar\footnote{$[X/H]={\rm log} [X/H]_{DLA}$-{\rm log}$[X/H]_{\odot}$}, 
and the ion used along that specific line of sight.
No dust depletion or ionization correction have been applied (see Sections \ref{sec:depl} and \ref{sec:ioncorr}).
In the last column we list the references relative to each GRB and, in the Appendix \ref{app:appA}, 
we briefly described each line of sight in more detail.


\section{Discussion}
\label{sec:discussion} 

\subsection{Possible Observational Biases}
\begin{figure*}[t!]
\epsscale{1.0}
\includegraphics[scale=0.55,angle=0]{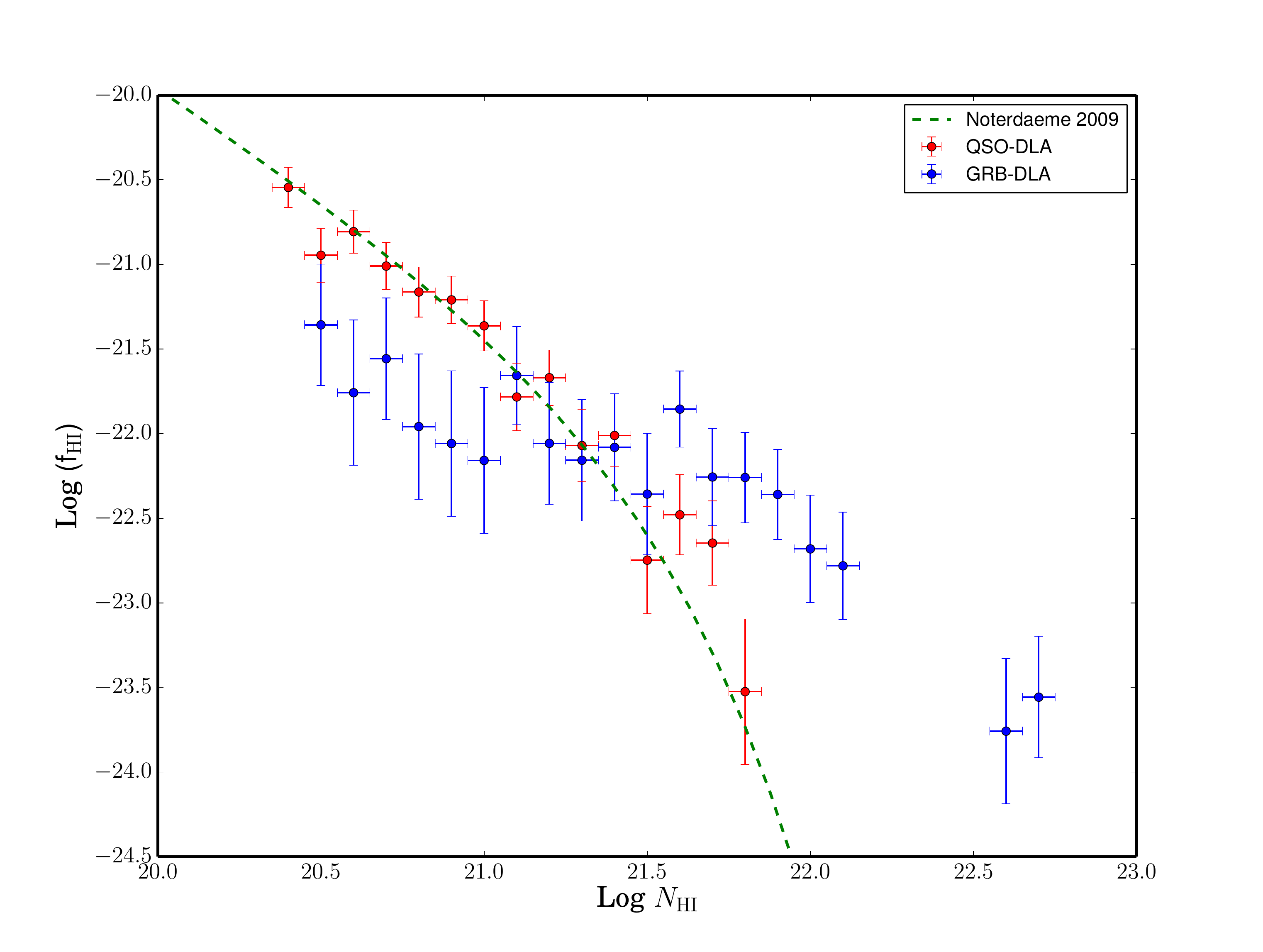}
\caption{\footnotesize{Column density distribution for our two DLA samples (QSO in red and GRB in blue). Vertical error bars are derived assuming poissonian
distribution (95\% confidence level) and we overplot in green (dashed line) the model
by \citet{Noterdaeme:2009aa}. 
Despite the fact that the GRB-DLA sample is a factor of 4 smaller there is a clear overlap in the distribution around $N_{\rm HI}$=21.5, while GRB-DLAs show a much larger number of dense systems compared with the QSO-DLA sample \cite[see also][]{Noterdaeme:2014aa}, indicating that GRBs are located, if not embedded, in very dense regions within their host galaxies, that can be the beacons of present star-formation.} 
}
\label{fig:hdistr}
\end{figure*}

First of all we need to understand if there are possible biases that might affect either samples:
GRBs are selected solely on the prompt emission detection in the gamma-ray energy bands
from dedicated satellites (e.g.\,\swift). The spectra were taken mostly independently of the brightness
of the afterglow, although a preference in observing brighter events might be present. Neither of these
two selection criteria (brightness in the gamma-ray bands or in the optical) seem to explain the difference in \nh\ distribution (Figure \ref{fig:hdistr}). 
As previously noted by  \citet{Pontzen:2010aa}, GRB-DLAs seem to probe a different type of absorbers than the QSO-DLA population.
In particular, if the faint afterglows were not observed spectroscopically because some high
level of extinction \citep[like the ``dark'' GRBs from][]{Perley:2013aa}, and assuming a possible correlation between metallicity and visual extinction \citep[see ][]{Zafar:2013aa}, this would bias us towards lower metallicities.
This, combined with a small number of metal-poor GRB-DLAs at \hiz\ suggests that, if such bias exists it would have a small effect, also because at high redshift there are less dusty systems \citep{Covino:2013aa}. At $z\gtrsim6$, our understanding of dust production mechanisms (e.g. SN, AGB stars) are still to be fully understood, although some advancements have been made both theoretically and observationally \citep{Gall:2011aa}.
We also note that, similarly to these authors, we are assuming the extinction along the line of sight as the same as the one derived from afterglow studies, which in principle can lead to different estimates with respect to the hosts galaxy extinction \citep[see][for a comparison between the afterglow- and host-derived extinction]{Elliott:2013aa,Perley:2013aa}. Overall, despite the large sample presented here, our conclusions, in particular at high redshift, 
require more data in order to better assess the effect of such observational biases.

The QSOs sample from R12 and R14, instead, has been selected solely on the presence of a \Lya\ line
in their SDSS spectrum and log$\nh\geq20.3$, therefore represent an unbiased sample with respect to the gas metallicity \citep{Rafelski:2012aa}. 
The $z\lesssim2$ QSO-DLAs were selected based on the presence of \MgII\ in the SDSS spectra, implying a small bias against low-metallicity systems at such redshift. 
Since our redshift range of interest is mostly at $z\gtrsim2$, this should not be a source of significant concern.

\subsection{Ionization Correction}
\label{sec:ioncorr}
The identification of fine-structure transitions and the line profile variability within the first hour after the GRB explosion
have been associated to the effect of an evolving radiation field from the GRB (like the afterglow) or nearby young stars onto the progenitor surrounding 
medium \citep{Prochaska:2006aa,Vreeswijk:2007aa,De-Cia:2012aa}. 
This processes directly affects metal abundance measurements and 
the inferred metallicity ionization correction is strongly dependent on the hydrogen column density \citep{Vreeswijk:2013aa}. 

Distinguishing between ionized (photoionized or excited) gas from the GRB emission in the vicinity of the GRB progenitor
and the bulk of the host galaxy ISM, which is also ionized by the surroundings radiation field, is a complicated process: 
if the gas probed by the afterglow spectroscopy
is close  to the GRB ($\lesssim 200$pc), then the ionization is increasing with time and complex modeling is 
required to estimate the time dependent ionization correction \citep{Krongold:2013aa}. 
Probably, the most important example in which a detailed photoionization modeling (including photo-ionization and excitation effects) has been performed, 
is the sub-DLAs GRB\,080310  \citep{Vreeswijk:2013aa}.
In most of our GRB-DLAs the data have been acquired at much later time and in several cases, when multiple spectra were obtained, these objects do not present line profile variation: such non-variability suggests that we are likely probing the ISM at larger distances and/or gas clouds unaffected by
the GRB radiation.

Reassuringly, similar models to the ones of \citet{Vreeswijk:2013aa} have been performed in GRB-DLAs spectra \citep[][]{Ledoux:2009aa}:
ionization corrections are often minimal ($\lesssim10\%$) compared to other even extreme cases, 
e.g. star-forming galaxies or Lyman-$\alpha$ emitters.
For these reasons we did not apply any ionization correction to our derived column densities and our final 
metallicity measurements. 
We stress that the ionization correction is an important aspect of the column density determination and requires particular attention and 
a much larger sample of rapid high-resolution spectroscopic observation sequences starting not later than a few minutes after the burst. 
While this is the least known quantity in our study we argue that our results are
indicative of an overall general characterization of the GRB-DLA population. A large sample of multi-epochs
high-resolution datasets and accurate modeling is still needed, but is beyond the scope of this paper.

\subsection{Depletion Correction}
\label{sec:depl}
Understanding the effect of dust depletion in DLAs requires the determination of ionic abundances of both
refractory and non-refractory elements \citep[see for example][]{DElia:2014aa}. 
\citet{Savaglio:2004aa} determined for the first time such ``depletion pattern'', while recently 
\citet{De-Cia:2013aa} compared a sample of 20 GRB-DLAs with 47 QSO-DLAs in order to 
study the dust-to-metal ratio of these systems. These authors used the [Fe/Zn] as dust
indicator and found that the dust depletion correction is in the most depleted cases $\sim+0.1$\,dex independently of metallicity.

These results are in contrast with R12 where  strong iron depletion is present at $Z/Z_{\odot}>-1.0$.
Instead, R12 shows that silicon is rarely depleted in QSO-DLAs and that depletion 
effects are relevant only at high metallicity ($\gtrsim-0.3\, Z/Z_{\odot}$), therefore are minimal for the majority of
our GRB lines of sights\citep[see also][]{Vladilo:2011aa}.

For our GRB-DLA sample, to calculate a depletion correction is not always possible, 
especially for our low resolution GRB-DLA spectra. 
Also, in the few cases where iron was used, we derived $Z/Z_{\odot}< -1.0$, so depletion correction should be minimal.

Nevertheless, the application of a depletion correction would increase the metallicity of our systems, strengthening some of our conclusions 
(see Section\,\ref{sec:conclusions}). Therefore, we choose not to apply any depletion correction to the measured metallicities.
Finally, we performed $\alpha$-enhancement
 correction to the iron based metallicities for our
high-resolution sample using the correction factors adopted by R12.

\subsection{Metallicity Evolution}
Our findings are shown in Figure \ref{fig:fits}, where the QSO-DLAs (in gray) and GRB-DLAs (in red)
metallicities are plotted. 
We perform a linear fit of the 
metallicity measurements (and limits) with redshift using a survival analysis technique \citep{Schmitt:1985aa},
which takes into account upper and lower limits. In particular we used the {\tt statistics.schmittbin} package within the {\tt IRAF}\footnote{IRAF is 
distributed by the National Optical Astronomy Observatory, which is operated by the Association of Universities for Research in Astronomy (AURA) under 
cooperative agreement with the National Science Foundation.} distribution. We also used a bootstrap sampling in order to determine the 1-sigma error in the fit (with 500 iterations).

Ideally, we would like to use our metallicity measurements to investigate the cosmic metal budget at different epochs: this is usually 
done weighting the average metallicity  over a specific redshift bin with the total neutral hydrogen column density in the same redshift interval
(see R12 and R14 for the QSO-DLAs sample).
Unfortunately, because the large number of limits present in the GRB-DLA sample we simply fit the metallicity of the single systems, 
which still provide useful insight on the DLA populations metal content.

For the GRB-DLA sample, we derive  $[X/H]_{GRB} = (-0.07\pm0.06) z - (0.75\pm0.25)$ (thick red dashed line), 
which is also consistent with no-evolution (at the $1\sigma$ confidence level).
For the QSO-DLA sample we derive a linear trend between redshift and DLA metallicity given by $[X/H]=(-0.20\pm 0.03) z-(0.68\pm0.09)$ (thick black dashed line in Figure \ref{fig:fits}).

To test the reliability of our fit and how much this is sensitive to the small number of abundance measurements
(with respecting the limits) we ran a series of Monte Carlo simulations: we created 1000 mock samples of GRB-DLAs metallicity measurements
and limits (with values within the typical GRB-DLA high-resolution points RMS) and we repeated the survival analysis fit.
Assuming an intrinsic true slope from the QSO-DLA distribution of $-0.2$ we obtained a slope 
 $>-0.07$ or less only in 7\% of the cases, which means that it is unlikely that our results are affected by 
low number statistics. Similar tests with different intrinsic distribution (from flat to very steep metallicity evolution)
provide similar results and 
reassure us that, despite the small number statistics, we can recover the input slopes to within the reported 1$\sigma$ 
confidence interval.

Finally, we point out that GRB-DLAs and QSO-DLAs have a similar metallicity distribution at $z\sim2-3$,
suggesting that there is no difference among the two populations of absorbers in terms of metal content. 
We performed a two-sample Kolmogorov-Smirnov test among these subsamples using the IRAF {\tt stsdas.analysis.statistics.twosampt} 
task and we can rule out the null hypothesis  that the two distributions are drawn from the same parent population
at the 90\% confidence level. Instead, if we naively consider the lower limits as actual measurements and apply an arbitrary
correction of different values (from +0.1 to +0.3 dex), there remains evidence that the two distributions are distinct 
even with the highest value considered (+0.3 dex), but the statistical significance is $<99\%$.

However, the GRB-DLA metallicity declines suggests that, in particular at $z\gtrsim3$, GRB-DLA environment 
is more metal enriched than in QSO-DLAs, likely by active star-formation episodes. These metals may have been ejected by supernova explosions or mass losses and polluted the GRB progenitor's
neighborhood before the GRB occurred \citep{Kroger:2006aa,Mao:2010aa,Matteucci:2001aa,Kulkarni:2013aa}. 
If this is true, the inferred metallicity may not reflect the overall metal content of 
such high-$z$ GRB host galaxies, although we know that the gas intercepted by GRB afterglow
spectra may lie up to few kpc from the GRB explosion site (see next section).

\subsection{Characterizing these DLA populations}
We will now try to understand the different metallicity  evolution between the GRB- and QSO-DLA samples.
For example, if GRB-DLAs trace the general host galaxies ISM, this gas ionization state may show the effect of 
whatever local ionization field is in the vicinity of the GRB or along the line of sight. 

\citet{Prochaska:2007ab} argued that the size of such molecular cloud would exceed the largest molecular cloud in the local group, 
so the GRB-DLAs are tracing material as far as at least 100pc 
from the GRB \citep[and even out to $\approx2$ kpc in the case of GRB\,060418,][]{Vreeswijk:2007aa}.
The fact that  in the $z=3.5-6$ redshift range the metallicity of the gas is, on average $10\% $ of the solar 
value suggests that a substantial amount of metals are already present at high redshift.

These estimates are in agreement
with the most recent GRB progenitor models \citep[][]{Woosley:2011aa,Woosley:2012aa}:
if the GRB host galaxy is metal rich, then a large amount of metals have been
produced by SNe or, in the highest redshift bin observed, by a late population of PopIII stars (or more likely early PopII). 
Nevertheless, in order to be able to produce a GRB explosion from Wolf-Rayet type stars, the amount of metal injected 
into the ISM and mixed throughout the host galaxy has to be below a certain  threshold to retain enough angular 
momentum and minimize the stellar mass loss of the progenitor \citep{Woosley:2006aa,Woosley:2011aa,Woosley:2012aa}.
Clearly more data need to be acquired, in particular at \hiz\, to confirm the existence of such limits,
also in light of the high-metallicity hosts observed \citep[e.g.][]{Elliott:2013aa,Kruhler:2012aa,Savaglio:2012ab}.

On the other hand, DLAs identified in QSO spectra are cross-section dependent,
and therefore they might not trace the denser part of the galaxies (like for GRB-DLAs).
Also, it has been proposed that a combined large sample of GRB- and QSO-DLAs
may represent a complete census of $z\approx3$ star-forming galaxies that could be missed 
by magnitude limited surveys \citep{Fynbo:2008aa}.

Finally, as R14 pointed out, the observed metallicity decrement at $z\gtrsim 4.7$ suggests 
an increase in the covering fraction of neutral gas: similar behavior in fact can be produced by the 
combination of increased density of the Universe and lower background radiation field, which
allows the hydrogen gas to be self-shielded at the lower density as function of redshift \citep{Fumagalli:2013aa}.
In this picture, the denser region would reside in the halo of the galaxies or in the IGM where
the star-formation, and therefore metal enrichment, is lower.

\subsection{GRB-DLA metallicity in context}

A long standing debate exists as to the degree to which GRBs faithfully trace the cosmic star formation rate.  Although GRBs have been associated with broad lined supernovae at low redshift and regions of active star formation in their host galaxies, spectroscopic observations have shown that GRB host galaxies tend to be relatively metal poor compared to SNe Ibc hosts \citep{Modjaz:2008aa}.  This observed preference for low metallicity environments may impart a redshift dependent bias in the type of star forming regions that can produce a GRB \citep{Kocevski:2009aa,Trenti:2014aa}.

In particular, at low redshifts ($z \lesssim1$), a preference for low metallicity environments would limit GRBs to low mass spirals and dwarf galaxies \citep[e.g. ][]{Levesque:2010aa}, due to the well established relationship between mass and metallicity \citep[see][ and references therein]{Kocevski:2011aa,Graham:2013aa}.  At higher redshifts, the mass range of galaxies capable of hosting a GRB would increase to include more massive, star forming galaxies, since the average metallicity of all galaxies in the Universe falls. Recent unbiased searches of GRB host galaxies like the \emph{THOUGH} survey \citep{Hjorth:2012aa} or of the host galaxies of the ``dark'' GRB population \citep{Perley:2013aa}  largely support this trend.  
These surveys find that bursts at intermediate redshifts tend to be drawn from star forming galaxies with a greater diversity of mass, morphology, and dust content, suggesting that high redshift GRBs may serve as more faithful tracers of cosmic star-formation compared to their low redshift counterparts \citep[see also][]{Hunt:2014aa,Perley:2014aa}.

Our sample of GRB-DLAs, although not a complete host sample, covers a much greater redshift range than the emission line derived measurements in these studies and, more importantly, does not depend on strong observational biases (e.g. brightness of the host, intensity of the emission lines), although we may 
miss some dusty (therefore metal rich) events.
Based on the arguments outlined above, even as metallicity biased tracers of star formation, the GRB-DLA results presented in Figure \ref{fig:fits}  should become more representative of the metallicity evolution of the general star forming galaxy population with increasing redshift. However, a better understanding of both observational biases  \citep{Fynbo:2008ab,Fynbo:2008aa} and the effect of metallicity in GRB production \citep{Jakobsson:2013aa} are required before we can fully address the connection between the environments which are capable of producing GRBs and the conditions of star forming regions in the early Universe. 

From a theoretical standpoint, recent simulations by \citet{Trenti:2014aa} have been suggesting that two combined channels of GRB 
populations exist: one, the ``collapsar'' mode, which strictly depends on the host metallicity, and a second one, the ``binary stars'' mode, which is metallicity independent.  
Assuming these two modes coexist and the known mass-metallicity relation of star forming galaxies, these authors 
predict GRB host galaxies metallicity redshift evolution with different combinations of these two channels (from strong metallicity bias to
an almost negligible one).
Unfortunately, metallicity measurements from emission line diagnostics are not yet available for $z\gtrsim1$ host galaxies, but
 our sample represents the best opportunity to test these models. 

In Figure \ref{fig:trenti} we present our metallicity results in comparison the predicted metallicity for the upper 95\%, 
the median, and the bottom 20\% of the GRB host galaxies distribution, assuming an absent (dotted lines) and a moderate (solid lines) 
metallicity bias \citep[adapted from][ using a value of $p=0.04$]{Trenti:2014aa}.

These models essentially predict that in any given redshift bin, for example, 20\% of the hosts that produce a GRB have metallicity below the lower solid line. Indeed, focusing on the redshift $z=2-3$ range, 5 over 27 hosts have metallicity below the line. On the other hand the ``no-metallicity bias'' model does not agree with the data.
Overall, it seems that at least 
at $z\lesssim4$ a moderate metallicity bias is required in order to reconcile theory and observations \citep{Vergani:2014aa}, although
a more detailed analysis of these models and the implications on the metallicity cut-off in the GRB host masses is needed,
in particular to understand the role of such metallicity bias, if indeed exists, at higher redshift. 

\section{Conclusions}
\label{sec:conclusions} 
In this work we investigated the properties of GRB lines of sight
that show evidence for  Damped Lyman-$\alpha$ systems within the GRBs host
galaxy. 
Similar systems have been studied along QSOs in order to understand the chemical 
enrichment and the metal evolution in cold gas systems over cosmic times.

We collected all the publicly available GRB afterglow spectra, including data already published 
in the literature and we uniformly analyzed these lines of sight in order to obtain metal
abundances using mildly or non refractory species (e.g. sulphur, silicon, and zinc).
In particular, we opted for a conservative approach when considering low-resolution data (see Section \ref{sec:resol}): we consider depletion correction to our metal abundance measurements a negligible effect (see Section \ref{sec:depl}).
Also, we note that ionization correction may have an important role in estimates derived from very early time spectroscopic data ($\lesssim 10$ minutes), due to the highly variable ionization flux from the GRB itself. Nevertheless, since most of our data have been acquired at much later times and the fact that detailed analyses are not always possible (especially with low-resolution data) we do not consider such correction in our analysis (see Section \ref{sec:ioncorr}).

\begin{figure*}[t!]
\epsscale{1.0}
\includegraphics[scale=0.5,angle=0]{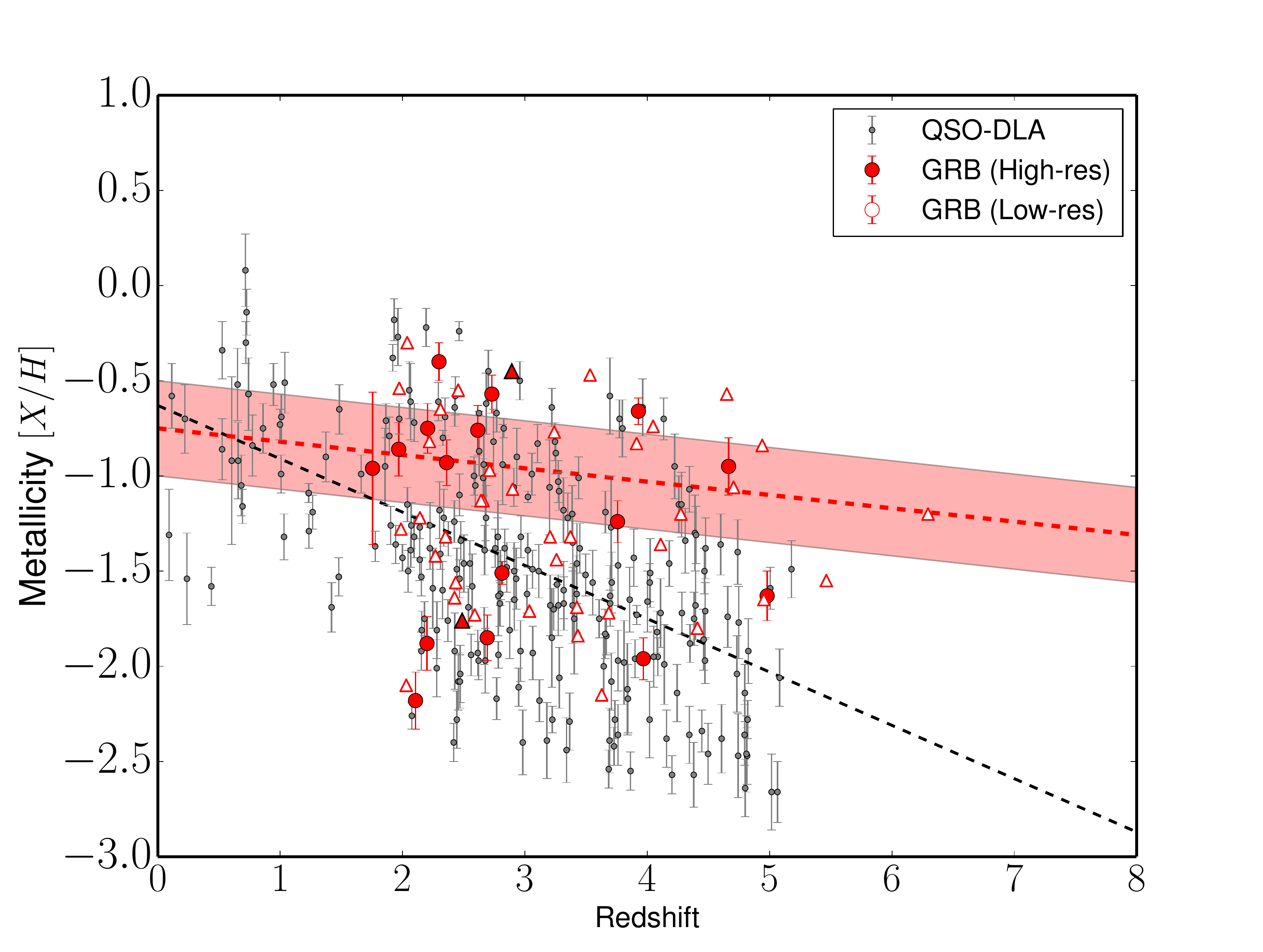}

\caption{\footnotesize Metallicity evolution with redshift for the GRB (red) and the QSO (grey) samples. Lower limits are indicated by upward triangles, while filled/open symbols indicate if these values come from high/low resolving power instruments. 
We perform a linear regression fit of the GRB-DLA data using
the Schmitt survival analysis method, which keeps into account the censoring within the dataset (red dashed line). 
The shaded area represents the $1\sigma$ error in the fitting parameters obtained using 500 bootstrap iterations.
A linear fit of the QSO-DLAs metallicity is marked by the dashed black line (see text for details). The GRB sample, despite the large scatter in [X/H], seems to probe an environment which slowly declines across the $z=1.8-6$ redshift range.} 
\label{fig:fits}
\end{figure*}

In Figure \ref{fig:fits} we present our GRB-DLAs metallicity evolution fit in comparison
with the QSO-DLA results from R12. We performed a detailed survival analysis fit to accurately take into account
all the limits present in our dataset. As in previous works \citep{Savaglio:2012ab,Arabsalmani:2014aa}
we derive a much shallower decline of the GRB host metallicity with redshift relative to 
the QSO-DLA sample  suggesting a somewhat metal rich environment for the GRB host galaxies (in particular
at $z>4$), but still below solar values, in agreement with 
GRB progenitor models \citep{Woosley:2012aa}.
At $z\lesssim 3$ there is a reasonable overlap between the two populations, according to their
metallicity properties, indicating similarities between the two DLA populations, despite the higher \HI\ column density
traced by the GRB-DLAs. 
At higher redshift, GRB-DLAs seem to prefer higher metallicities than QSO-DLAs.
Despite the small number of high-resolution data, a few lines of sight (e.g. GRB130606A or GRB060510B)
seem to indicate that these GRB-DLAs point towards a denser, metal rich environment,
likely tracing a less common population of metal-rich DLAs.

Overall, our findings confirm the idea that moving towards higher redshifts, 
GRBs trace preferentially a denser metal rich environment within  galaxies, 
while the QSO-DLA population may be progressively dominated by neutral regions with minor star formation. 
In other words, the absence of metal poor $([X/H]_{GRB}\lesssim -1.5)$ GRB-DLAs at \hiz\ seems to indicate that
such GRB-DLA hosts are at the low-end of the luminosity (or mass) distribution and do not present
 high star-formation (and GRB progenitors). 
Another possibility is that these GRBs occurred in a completely different environments 
(galactic haloes). Deep imaging, with 8-10 meter telescopes or with \hst, is needed in order to
 fully characterize these hosts.

\begin{figure*}[t!] 
\epsscale{1.0}
\includegraphics[scale=0.60,angle=0]{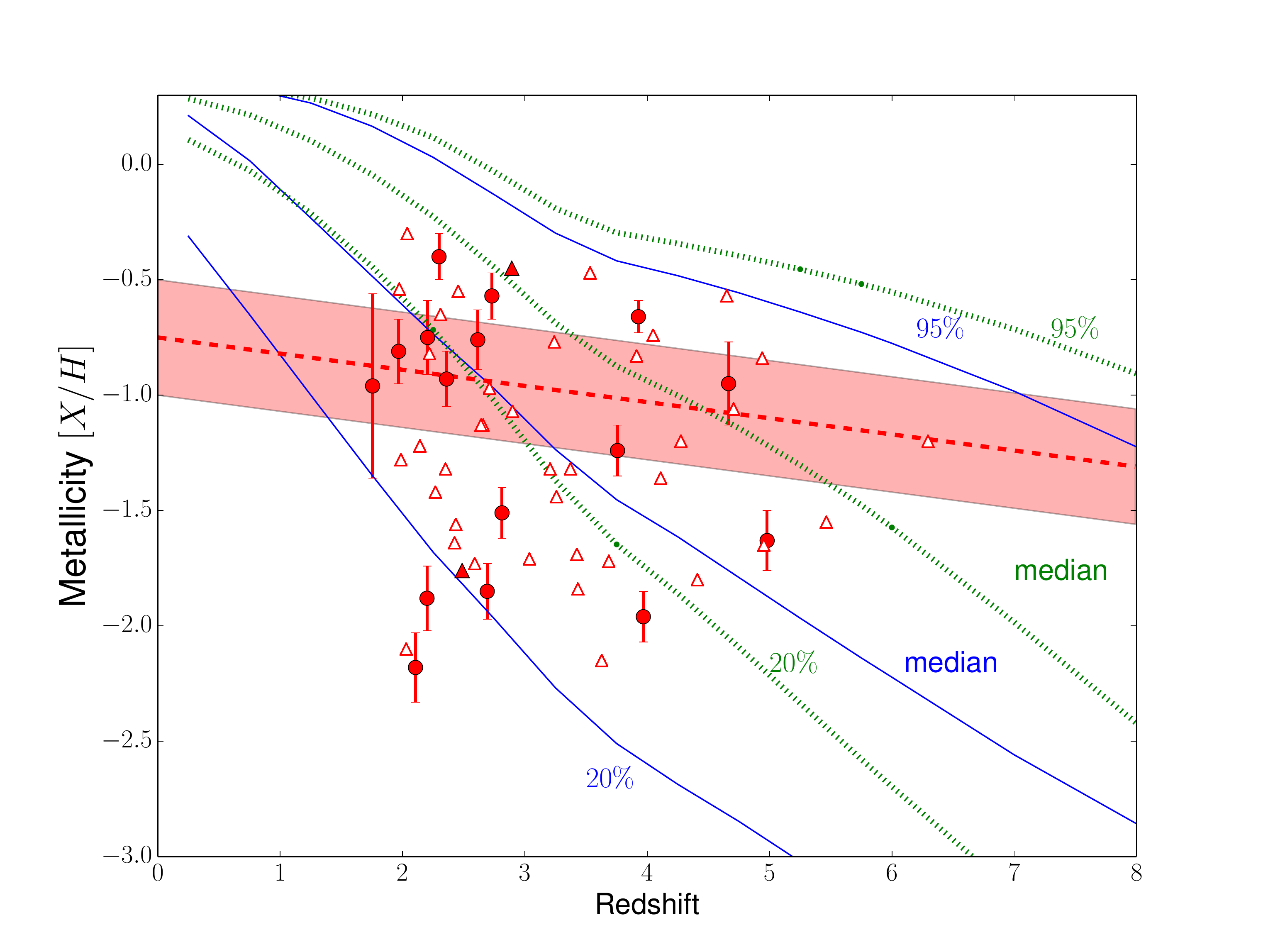}
\caption{\footnotesize{Adapted from \citet{Trenti:2014aa}. The colored lines represent two particular model for GRB hosts metallicity assuming a mild metallicity bias ($p=0.04$, solid blue) and no bias at all (dotted green). Lines also indicate the trends for the top 95\%, the median, and lower 20\% of the simulated GRB host population. The mild bias model is broadly capable to reproduce the observed metallicity distribution of our biased sample of GRB-DLAs at a specific redshift, while the ``no bias'' is largely inconsistence with the data.}}
\label{fig:trenti}
\end{figure*}

Finally, we compare our findings with the most recent model prediction of GRB host galaxy metallicities 
\citep[Figure \ref{fig:trenti},][]{Trenti:2014aa}. Our results broadly agree with host galaxy metallicity predictions
 where two-channels for GRB productions are considered (collapsars and binary) and mildly affected
by a metallicity bias.
The model predicts well the metallicity distribution for the 
bottom 20\% and median of the host population, though more work is required to fully understand this
metallicity bias and its effect on other properties of the GRB host population \citep[e.g. the hosts mass limit driven by
a metallicity cut-off][]{Kocevski:2009aa}.

Our GRB-DLA host galaxies represent the largest sample available to date and, although not complete, 
it is suitable for multi band follow-up, 
in particular for the current and upcoming near-infrared spectroscopic
instruments, which will allow to determine SFR and metallicities directly using emission line diagnostics.
Also, the determination of the hosts properties, like  mass and rest-frame UV star formation, will help in 
better characterizing the overall \hiz\ GRB host population, and their capability to harbor GRB progenitors.
Furthermore, we will be able to better understand the observational biases that might affect our results, especially at 
\hiz, where few lines of sights are observed (including dust extinction and/or cosmic metallicity trend). 

While multiband surveys of this sample of GRB-DLA hosts will allow a better characterization of these galaxies,
the advent of future missions like \emph{JWST} and the new generation of 30-m telescopes will be able to
 identify these faint hosts at the highest redshifts and spatially resolve the regions of star-formation 
 traced by GRB-DLAs, which seem to hide the secrets of primordial star formation sites.

\acknowledgments
This research was supported by the NASA Postdoctoral Program at the Goddard Space Flight Center, 
administered by Oak Ridge Associated Universities through a contract with NASA.
A.C. thanks M.F. and M.R. for the incredible support and useful discussions during the
writing of this work. A.C. also thanks M.T. for fundamental discussion as well for providing 
the models curves.
MF acknowledges support by the Science and Technology Facilities Council (Grant \#ST/L00075X/1).
A.C. thanks P. D'Avanzo for providing the spectrum of GRB 090205A.

\clearpage

\input{tab1_paper_ref.tex}

\clearpage

\clearpage

\clearpage


\clearpage


\bibliographystyle{apj_8}
\bibliography{grbdla}

\appendix 
\section{Description of relevant spectra}
\label{app:appA}
\subsection{GRB000926}
Our results confirm the finding published in the literature. The low-resolution data though allow to place only a lower limit
based on the \ZnII\ lines \citep[also consistent with ][]{Prochaska:2006ab}.
\subsection{GRB011211}
Our examination of the FORS spectrum allows to place a lower limit based on the \SiII\ line of $[\rm{X/H}]\gtrsim -1.22$, consistent with the result of \citep[][]{Prochaska:2006ab}.
\subsection{GRB020124}
We derived similar \nh value to the published work of \cite{Hjorth:2003aa}, but the spectrum is too low S/N to provide an adequate metallicity measurement.
\subsection{GRB021004}
Extensive literature is present for this GRB sub-DLA \citep[e.g. ][]{Fiore:2005aa,Fynbo:2005aa,Lazzati:2006aa,Castro-Tirado:2010aa}. We report this GRB here for completeness.
\subsection{GRB030226}
We also notice the presence of an intervening system along this sightline at $z_{int}=1.96236$. Based on some isolated 
Iron lines we place a lower limit on this GRB-DLA consistent with the measurement of \citep{Shin:2006aa}. Our iron-derived metallicity limit is consistent also with the measurement reported by \citep{Schady:2011aa}, considering the large uncertainty in the \HI\ estimate (0.3 dex).
\subsection{GRB030323}
Our analysis is consistent with the previous work by \citet{Vreeswijk:2004aa} (including the different in the sulphur solar abundance between  
\cite{Asplund:2009aa} and \cite{Grevesse:1998aa}) and \citet{Prochaska:2006ab}.
\subsection{GRB030429}
The spectra has been presented before \cite{Jakobsson:2004aa}. The lower limit in the metallicity is given by the few detected lines (e.g. \SiII).
\subsection{GRB050319}
The spectrum has been presented in \cite{Fynbo:2009aa}, we adopt the measurement from \cite{Laskar:2011aa} as
an upper limit based on \SiII\,1526.
\subsection{GRB050401}
The spectrum was presented by \cite{Watson:2006aa}. We obtain similar results and adopt a value of $[\rm{X/H}]\gtrsim -1.07$ based on \ZnII\ line measurement.
\subsection{GRB050505}
The metallicity we derived from sulfur measurements is similar to what has been presented by \cite{Berger:2006aa},
though we consider this to be a lower limit.
\subsection{GRB050730}
The high-resolution allows accurate determination of metallicity using different tracers. Our results presented in Table \ref{tab:list} are consisting with the published values \cite{DElia:2007aa} and \cite{Prochaska:2008aa}.
\subsection{GRB050820}
The measurement we obtained based on sulfur lines is consistent with the results from \citet{Jakobsson:2006aa} and \citet{Prochaska:2007ab}.
\subsection{GRB050904}
As noted by \cite{Thone:2013aa}, the values reported in the literature \citep{Kawai:2006aa} is likely an overestimation due to blending of different lines, so we adopt their sulphur measurement, $[\rm{S/H}]\gtrsim -1.0$, as a lower limit for this GRB.
\subsection{GRB050908}
This is a sub-DLA system, so it is included for completeness.
\subsection{GRB050922C}
The values reported by \citet{Fox:2008aa} and  \citet{Prochaska:2007ab} are consistent with our own measurement.
\subsection{GRB051111}
Only upper limit has been derived for this GRB.
\subsection{GRB060115}
These date are part of the compilation from \cite{Fynbo:2009aa}. Our estimated lower limit is mostly based on \SII\ lines.
\subsection{GRB060124}
This is a sub-DLA system, so it is included for completeness.
\subsection{GRB060206}
Our analysis is consistent with the results of \cite{Fynbo:2006aa}, but
we consider our estimated value of $[\rm{X/H}]= -0.74$ a lower limit, based on the strength of the same sulphur lines.
\subsection{GRB060210}
This GRB spectrum shows the presence of a nearby intervening systems at $z_{int}=3.817$. Despite its presence
a stringent lower limit of $[\rm{X/H}]\geq -0.83$ can be placed based on iron and silicon lines.
\subsection{GRB060223}
The value from \cite{Chary:2007aa} has been adopted.
\subsection{GRB060510B}
We confirm the result from the literature based on sulfur lines \cite{Chary:2007aa,Price:2007aa}.
\subsection{GRB060522}
The value of \nh\ from \cite{Chary:2007aa} has been adopted, but there are no metal lines detected.
\subsection{GRB060526}
This is a sub-DLA system, so is included for completeness.
\subsection{GRB060605}
This is a sub-DLA system, so is included for completeness.
\subsection{GRB060607A}
This is a sub-DLA system, so is included for completeness.
\subsection{GRB060707}
We confirm the analysis of \cite{Jakobsson:2006aa}, but the low-resolution data
allows only to place a conservative lower limit of $[\rm{X/H}]\geq -1.69$ based on weak \FeII\
transitions.
\subsection{GRB060714}
The spectrum has a higher $S/N$ and similarly to GRB060707 we could derive a lower limit, 
but in this case we used \ZnII\ lines \citep{Jakobsson:2006aa}.
\subsection{GRB060906}
This spectrum was presented by  \cite{Jakobsson:2006aa} and metallicity measurements were recently 
presented by \cite{Laskar:2011aa}. Using Sulfur lines we derive $[\rm{X/H}]\geq-1.72$.
\subsection{GRB060926}
For this GRB we used likely saturated \ZnII\ lines and place a lower limit of $[\rm{X/H}]\geq-1.32$.
\subsection{GRB060927}
This low $S/N$ spectrum has been published before \citep{Ruiz-Velasco:2007aa}. We derived more stringent limit
then the ones derived by \citet{Laskar:2011aa}  using \SII\ saturated lines.
\subsection{GRB061110B}
This spectrum is part of the compilation of \cite{Fynbo:2009aa}. 
We obtained similar measurement than \cite{Laskar:2011aa}, but we consider this a lower limit.
\subsection{GRB070110}
This spectrum is part of the compilation of \cite{Fynbo:2009aa}, we performed our own metallicity measurement.
\subsection{GRB070411}
This is a sub-DLA system, so is included for completeness.
\subsection{GRB070506}
This spectrum is part of the compilation of \cite{Fynbo:2009aa}. 
We also identified an intervening system at $z=2.071$. We derived a metallicity lower limit
 for the host from \ZnII\ lines.
\subsection{GRB070721B}
This spectrum is part of the compilation of \cite{Fynbo:2009aa}. 
The low-resolution of this spectrum allows only a lower limit estimate from \SiII.
\subsection{GRB070802}
This spectrum is part of the compilation of \cite{Fynbo:2009aa} and has been studied by \cite{Eliasdottir:2009aa}. 
The low-resolution of this spectrum allows only a lower limit estimate from blended \SiII: we are a bit more conservative than
a previous work by \cite{Eliasdottir:2009aa}, but the two values are within 0.1 dex, which is below the typical uncertainty for these
low-resolution datasets.
\subsection{GRB071020}
This GRB is likely a sub-DLA system, so is included for completeness.
\subsection{GRB071031}
This spectrum was presented by \cite{Ledoux:2009aa}.
Most of the useful metal lines are saturated (e.g. \SII\ 1259).
Using weak \FeII\ features (e.g. \FeII\ 1611) we obtained similar results
$[\rm{X/H}]=-1.85\pm 0.12$ \citep[see also ][]{Fox:2008aa}.
\subsection{GRB080210}
This GRB was observed with the  FORS2 instrument at several different 
resolutions \citep[see ][]{De-Cia:2011aa}. Our results, based on our analysis of the same FORS2 data
shows similar results, though we consider these metallicity estimate a lower limit (see Section \ref{sec:resol}).
\subsection{GRB080310}
This GRB is likely a sub-DLA system, so is included for completeness.
\subsection{GRB080413A}
This spectrum was presented by \cite{Fynbo:2009aa} and metallicity estimated by \cite{Ledoux:2009aa}. We derived only a lower limit based on zinc and nickel lines, which are also consistent with these previous works.
\subsection{GRB080607}
This GRB spectrum was presented by \cite{Prochaska:2009aa} and it shows for the first time the clear presence in a GRB afterglow spectrum of \H2\ molecular lines. Using different metallicity tracers, we retrieve similar values for the metallicity limits ($[\rm{X/H}]\geq-1.71$).
\subsection{GRB080721}
This spectrum was presented by \cite{Starling:2009aa}. We derived a more 
conservative lower limit, using multiple $\alpha$-elements lines.
\subsection{GRB080804}
The spectrum was presented by \cite{Fynbo:2009aa}. We determine
the metallicity and relative depletion using Iron as well as non-refractory 
elements.
\subsection{GRB080810}
This GRB is likely a sub-DLA system, so is included for completeness.
\subsection{GRB080913}
This GRB is likely a sub-DLA system, so is included for completeness.
\subsection{GRB081008}
This spectrum has been presented by \cite{DElia:2011aa}. We obtained 
similar metallicity estimate. We also analyzed a Gemini/GMOS spectrum from which we derived consistent lower limits.
We used our UVES spectrum measurements.
\subsection{GRB090205}
This line of sight appears in \cite{DAvanzo:2010aa}.
In particular, we note that these author reported sulfur column abundance from blended profiles.
We derived a similar metallicity lower limit of $[\rm{X/H}]\geq-0.57$.
\subsection{GRB090323}
This GRB has a complex structure, with two absorbers within 600\kms\ \cite{Savaglio:2012ab}, one of which is a sub-DLA. Due to the complexity of this line of sight and the low-resolution of the VLT/FORS spectrum, we exclude GRB 090323 from our analysis. We report the metallicity in Table \ref{tab:list} for completeness.
\subsection{GRB090426}
This GRB is likely a sub-DLA system, so is included for completeness.
\subsection{GRB090516}
These data were presented by \cite{de-Ugarte-Postigo:2012aa}, but only equivalent widths were provided. Our analysis of the low-resolution VLT/FORS2 spectrum
results in a lower limit of the metallicity of $[\rm{X/H}]\geq-1.36$.
\subsection{GRB090809}
We measured a slightly higher metallicity than the one obtained by \cite{skula10}: nevertheless our silicon-derived metallicity value 
(see Table \ref{tab:list}) is well within 2$\sigma$ from the one presented by these authors. 
\subsection{GRB090812}
These data were presented by \cite{de-Ugarte-Postigo:2012aa}. We estimated a lower limit based on alpha-elements absorption features.
\subsection{GRB090926}
A VLT/FORS2 spectrum was presented by \cite{Rau:2010aa}, while an X-Shooter
series of spectroscopic observations was presented by \cite{DElia:2010aa}. 
We obtained similar results, in particular the X-Shooter resolution allows stronger constrain on the metallicity.
\subsection{GRB100219}
This spectrum was presented by \cite{Thone:2013aa}. We supplement this
dataset with a Gemini/GMOS spectrum. We also reanalyzed the X-Shooter spectrum.
Our measurement is slightly higher then the one presented by these authors, but within their error (0.2 dex).
\subsection{GRB100425}
We analyzed this dataset using the phase 3 products provided by the ESO database. 
Our metallicity lower limit is consistent with the one derived by \cite{skula10}.
\subsection{GRB110205}
This spectrum was already presented by \cite{Cucchiara:2011ab}. 
\subsection{GRB111008A}
This X-Shooter spectrum was presented by \cite{Sparre:2014aa}. Our Iron and sulfur metallicity estimate are consistent with
 the ones presented in this work. We report the former as metallicity tracer.
\subsection{GRB111107A}
This X-Shooter spectrum has not been presented in previous work. The signal-to-noise of the spectra is low, and therefore our metallicity measurement, based on 
saturated sulphur lines, has to be considered a lower limit.
\subsection{GRB120327A}
This spectrum has been published by \cite{DElia:2014aa}. Our sulphur measurement agrees with these authors.
\subsection{GRB120716A}
This spectrum has not been published in the literature before.
The spectrum was obtained 2.6 days after the GRB has been discovered. Despite the 
low $S/N$ we report a metallicity of $[\rm{X/H}]\geq-1.76 $ based on identified iron lines.
\subsection{GRB120815A}
This GRB spectrum has been published by \cite{Kruhler:2013aa}. Our analysis
is consistent with these authors results.
\subsection{GRB120909A}
This X-Shooter spectrum has not been presented before. We derive metallicity measurements from weak Iron lines as well $\alpha$-elements.
\subsection{GRB121024A}
This X-Shooter spectrum has been recently presente by \cite{Friis:2014aa}. 
We also identified multiple systems at $z_1=2.3014$ and an intervening one at $z_2=2.2977$
(corresponding to 400 \kms). Fine-structure transitions are identified in correspondence of $z_2$ system, suggesting that the cloud at $z_1$ is at large distance from the GRB radiation field. The broad, saturated \Lya\ profiles makes hard to discern between the two components. Therefore we opted for considering them as one single absorber. The metallicity from zinc lines is reported in Table \ref{tab:list}, while from iron lines we can infer
a large depletion factor \cite[see ][for a detailed analysis]{Friis:2014aa}.
\subsection{GRB121201A}
This X-Shooter spectrum was preliminary presented in GCN only \citep{Sanchez-Ramirez:2012aa}.
This line of sight present a possible \Lya\ emission line.
Unfortunately the signal to noise is low (S/N$\lesssim3$) and at the line redshift ($z=3.385$) it is difficult to identify metal lines unambiguously.
We therefore decide to report only the neutral hydrogen column density.
\subsection{GRB130408A}
This GRB has been observed by our team \cite{Tanvir:2013aa} as well as by VLT/X-Shooter \cite{Hjorth:2013aa}. We analyzed both spectra and report the metallicity measurements from both analysis. The X-shooter derived value is plotted in Figure\,\ref{fig:fits}.
\subsection{GRB130505A}
We present our Gemini spectrum for the first time. We identified several absorption 
lines, and we were able to determine a lower limit on the metallicity based on iron weak transition (\FeII1608).
\subsection{GRB130606A}
This spectrum has been published by \cite{Chornock:2013aa}.
The line of sight present signature of a likely sub-DLA system, therefore we present this value here for completeness.
\subsection{GRB140226A}
This spectrum was obtained by the Keck/LRIS instrument. Several metal lines have been identified, but the low-resolution of this instrument allows to place only a metallicity lower limit from Iron and Sulphur lines.
\subsection{GRB140311A}
This spectrum was obtained by our collaboration and present few metal absorption features. We were only able to place a lower limit on the metallicity based on nickel lines, since sulphur lines seems contaminated by other lines.
\subsection{GRB140419A}
This is a sub-DLA system, so is included for completeness.
\subsection{GRB140423A}
We present our analysis on our Gemini/GMOS spectrum \citep{Tanvir:2014aa}.
Based on $\alpha$-elements lines and weak iron lines (e.g. \FeII\ 1608) we were
able to place a lower limit on the metallicity.
\subsection{GRB140518A}
This data have not been published before, but preliminary analysis appears in \cite{Chornock:2014aa}.
Our metallicity limit comes from saturated sulphur transitions.

\end{document}

%% file: defs.txt
\def\hst{\emph{HST}}

\def\CIVdblt{{\rm C~}\kern 0.1em{\sc iv}~$\lambda\lambda 1548, 1550$}
\def\MgIIdblt{{\rm Mg~}\kern 0.1em{\sc ii}~$\lambda\lambda 2796, 2803$}
\def\NVdblt{{\rm N}\kern 0.1em{\sc v}~$\lambda\lambda 1238, 1242$}  
\def\OVIdblt{{\rm O}\kern 0.1em{\sc vi}~$\lambda\lambda 1031, 1037$}
\def\SiIVdblt{{\rm Si~}\kern 0.1em{\sc iv}~$\lambda\lambda1394, 1403$}
\def\AlIIIdblt{{\rm Al~}\kern 0.1em{\sc iii}~$\lambda\lambda1855,1863$}
\def\FeIIdblt{{\rm Fe~}\kern 0.1em{\sc ii}~$\lambda\lambda 2383, 2600$}

\def\AlII{\hbox{{\rm Al~}\kern 0.1em{\sc ii}}}
\def\AlIII{\hbox{{\rm Al~}\kern 0.1em{\sc iii}}}
\def\CaI{\hbox{{\rm Ca}\kern 0.1em{\sc i}}}
\def\CaII{\hbox{{\rm Ca}\kern 0.1em{\sc ii}}}

\def\CrII{\hbox{{\rm Cr~}\kern 0.1em{\sc ii}}}
\def\CI{\hbox{{\rm C~}\kern 0.1em{\sc i}}}
\def\CII{\hbox{{\rm C~}\kern 0.1em{\sc ii}}}
\def\CIII{\hbox{{\rm C~}\kern 0.1em{\sc iii}}}
\def\CIV{\hbox{{\rm C~}\kern 0.1em{\sc iv}}}
\def\CV{\hbox{{\rm C}\kern 0.1em{\sc v}}}

\def\H{\hbox{{\rm H}}}
\def\HI{\hbox{{\rm HI}}}
\def\H2{\hbox{{\rm H$_2$}}}

\def\HII{\hbox{{\rm H}\kern 0.1em{\sc ii}}}

\def\Lya{\hbox{{\rm Ly}\kern 0.1em$\alpha$}}

\def\Lyb{\hbox{{\rm Ly}\kern 0.1em$\beta$}}
\def\Lyg{\hbox{{\rm Ly}\kern 0.1em$\gamma$}}
\def\Lyfive{\hbox{{\rm Ly}\kern 0.1em$5$}}
\def\Lysix{\hbox{{\rm Ly}\kern 0.1em$6$}}
\def\Lyseven{\hbox{{\rm Ly}\kern 0.1em$7$}}
\def\Lyeight{\hbox{{\rm Ly}\kern 0.1em$8$}}
\def\Lynine{\hbox{{\rm Ly}\kern 0.1em$9$}}
\def\Lyten{\hbox{{\rm Ly}\kern 0.1em$10$}}
\def\HeI{\hbox{{\rm He}\kern 0.1em{\sc i}}}
\def\HeII{\hbox{{\rm He}\kern 0.1em{\sc ii}}}
\def\FeI{\hbox{{\rm Fe~}\kern 0.1em{\sc i}}}
\def\FeII{\hbox{{\rm Fe~}\kern 0.1em{\sc ii}}}
\def\FeIII{\hbox{{\rm Fe~}\kern 0.1em{\sc iii}}}
\def\MnII{\hbox{{\rm Mn}\kern 0.1em{\sc ii}}}
\def\MgI{\hbox{{\rm Mg~}\kern 0.1em{\sc i}}}
\def\MgII{\hbox{{\rm Mg~}\kern 0.1em{\sc ii}}}
\def\MgIII{\hbox{{\rm Mg~}\kern 0.1em{\sc iii}}}
\def\MgIV{\hbox{{\rm Mg~}\kern 0.1em{\sc iv}}}
\def\NaI{\hbox{{\rm Na}\kern 0.1em{\sc i}}}
\def\NV{\hbox{{\rm N}\kern 0.1em{\sc v}}}
\def\NII{\hbox{{\rm N}\kern 0.1em{\sc ii}}}
\def\NIII{\hbox{{\rm N}\kern 0.1em{\sc iii}}}
\def\NiI{\hbox{{\rm Ni~}\kern 0.1em{\sc i}}}
\def\NiII{\hbox{{\rm Ni~}\kern 0.1em{\sc ii}}}
\def\OVI{\hbox{{\rm O}\kern 0.1em{\sc vi}}}
\def\OI{\hbox{{\rm O}\kern 0.1em{\sc i}}}
\def\OII{\hbox{[{\rm O}\kern 0.1em{\sc ii}]}}
\def\SiII{\hbox{{\rm Si~}\kern 0.1em{\sc ii}}}
\def\SiIII{\hbox{{\rm Si~}\kern 0.1em{\sc iii}}}
\def\SiIV{\hbox{{\rm Si~}\kern 0.1em{\sc iv}}}
\def\SII{\hbox{{\rm S~}\kern 0.1em{\sc ii}}}
\def\SIII{\hbox{{\rm S~}\kern 0.1em{\sc iii}}}
\def\SIV{\hbox{{\rm S~}\kern 0.1em{\sc iv}}}
\def\TiII{\hbox{{\rm Ti}\kern 0.1em{\sc ii}}}
\def\ZnII{\hbox{{\rm Zn~}\kern 0.1em{\sc ii}}}

\newcommand{\kms}{\hbox{km~s$^{-1}$}}
\newcommand{\cmsq}{\hbox{cm$^{-2}$}}

\def\kms{\hbox{km~s$^{-1}$}}      
\def\cmsq{\hbox{cm$^{-2}$}}
\def\swift{\emph{Swift \,}}

\def\swift{\emph{Swift}}

\def\simlt{\mathrel{\hbox{\rlap{\hbox{\lower4pt\hbox{$\sim$}}}\hbox{$<$}}}}
\def\simgt{\mathrel{\hbox{\rlap{\hbox{\lower4pt\hbox{$\sim$}}}\hbox{$>$}}}}
\newcommand{\hiz}{high-$z$}

\newcommand{\nh}{\mbox{$N_{\rm HI}$}} 
\newcommand{\NH}{\mbox{$N_{\rm HI}$}} 

\newcommand{\NHunits}{\mbox{$\times 10^{20}\, {\rm cm}^{-2}$}}

%% file: tab1_paper_ref.tex
\begin{deluxetable}{lllllcllll}
\tablewidth{0in}
\tablecaption{GRB-DLAs Sample\label{tab:list}}
\tabletypesize{\footnotesize} 
\tablehead{
\colhead{GRB} & \colhead{$z_{GRB}$} &  \colhead{{\rm log}(\nh)} &
\colhead{$[\rm{X/H}]^a$} & \colhead{Ion}& Fine-&  \colhead{Telescope/}& \colhead{Resolution}& \colhead{S/N}&Reference\\
	&&&&&Structure&\colhead{Instrument}&\colhead{at 6000\AA}&per pixel&
}
\startdata
000926	&	2.3621	&$	21.3	\pm	0.25	$	& $\geq -0.30$	& Zn& N&Keck/ESI&20000&10&	[1]\\
011211	&	2.1427&	$ 20.4	\pm	0.2	$	& $\geq-1.22$	&	Si&N &VLT/FORS2 &2400&10&[2]\\
020124	&	3.198	&$	21.7	\pm	0.2	$	&	...	&...	&N&VLT/FORS1&450&4&[3]\\
021004	&	2.3289	&$	19.0	\pm	0.2	$	&	...&	... &Y &VLT/UVES&40000&6&[4][5]\\
030226	&	1.98	&	$ 20.5	\pm	0.3	$	&	$\geq-1.28$& Fe&Y&Keck/ESI&20000&40&	[6]\\
030323	&	3.3714	&$	21.9	\pm	0.07	$	&	$\geq-1.32$& S&Y&VLT/FORS2&2100&20&	[7]\\
030429	&	2.658	&$	21.6	\pm	0.2	$	&	$\geq -1.13$	&	Si	&  N&VLT/FORS1&600&40&[8]\\
050319	&	3.24	&	$ 20.9	\pm	0.2	$	&	$>-0.77{^c}$&S&N&NOT/AlFOSC&355&4&[9][10]\\
050401	&	2.899	&$	22.6	\pm	0.3	$	&	$\geq-1.07$	&	Zn & Y&VLT/FORS2&545&10&[11]\\
050505	&	4.27	&	$ 22.05	\pm	0.1	$	& $\geq -1.2 $		& S&	Y&Keck/LRIS&1200&20& [12]\\
050730	&	3.96723	&$	22.1	\pm	0.1	$	&$	-1.96\pm0.11$	&S	&Y&VLT/UVES&40000&10&[13][14]\\
050820A	&	2.6145	&$	21.1	\pm	0.1	$	&	$-0.76\pm 0.13$&S	&Y&VLT/UVES&40000&12&[8]\\
		&	 		& 					&	$-0.78\pm 0.11$&Fe	&Y&Keck/HIRES&30000&10&[14]\\
050904	&	6.26	&	$ 21.3	\pm	0.2	$	&$	\geq-1.0 {^c}$	&S	&Y&Subaru/FOCAS&1000&7&	[15]\\
050908	&	3.344	&$	19.4	\pm	0.2	$	&  ...	&...	&N&Gemini/GMOS&1200&20&[16]\\
050922C	&	2.1996	&$	21.55	\pm	0.1 $		&$-1.88\pm 0.14$&S&Y&VLT/UVES&45000&10&[17]\\
051111	&	1.549	&$	<21.9	$	&	...	&...&Y&Keck/HIRES&55000&20&[18]\\
060115	&	3.533	&$	21.5	\pm	0.1	$	&	$>-1.53$	&S&Y&VLT/FORS1&990&4&[9]\\
060124	&	2.3	&	$ 18.5	\pm	0.5	$	&	...& ...&N&Keck/LRIS&1200&18&[9]\\
060206	&	4.048	&$	20.85	\pm	0.1 $		&$\geq-0.74$	&S&Y&Lick/KAST&1200&28&[9]\\
060210	&	3.913	&$	21.55	\pm	0.15	$	&$\geq-0.83 $	&Si&Y&Gemini/GMOS&1200&$40$&[9]\\
060223A	&	4.41	&	$21.6	\pm	0.1		$&	$>-1.8^{c}$&S&N&	Keck/LRIS&1200&...&[19]\\
060510B	&	4.94	&	$21.3	\pm	0.1	$	&$	\geq-0.84$	& S&	N&Gemini/GMOS&1200&15&[19]\\
060522	&	5.11	&	$21.0	\pm	0.3^{c}	$&	...&...&N&Keck/LRIS&1200&2&[19]\\
060526	&	3.221	&$	19.9	\pm	0.15	$	&	...&...&N&VLT/FORS1&1200$^{b}$&18&[20]\\
060605	&	3.773	&$	18.9	\pm	0.4	$	&	...	&...&Y&PMAS&500&7&[21]\\
060607A	&	3.075	&$	16.95	\pm	0.03 $	&	...	&...&Y&VLT/UVES&55000&30&[17]\\
060707	&	3.425	&$	21.0	\pm	0.2		$&	$\geq -1.69$	&Fe&Y&VLT/FORS2&800&7&[22]\\
060714	&	2.711	&$	21.8	\pm	0.1	$	&	$\geq-0.97$	&Zn&Y&VLT/FORS1&800&30&[22]\\
060906	&	3.686	&$	21.85	\pm	0.10 $		&	$	\geq-1.72$	&S&N&VLT/FORS1&800&8&[22]\\
060926	&	3.206	&$	22.6	\pm	0.15	$	&	$\geq-1.32$	&Zn	&Y&VLT/FORS1&800&20&[22]\\
060927	&	5.464	&$	22.50	\pm	0.15	$	&	$\geq-1.55$	&S&N&VLT/FORS1&500&3&[23]\\
061110B	&	3.433	&$	22.35	\pm	0.10 $		&	$\geq-1.84$	&S&Y&VLT/FORS1&800&10&[9]\\
070110	&	2.351	&$	21.7	\pm	0.1		$&	$\geq-1.32$&Si	 &Y&VLT/FORS2&800&15&[9]\\
070411	&	2.954	&$	19.3	\pm	0.3	$	&...	&...&Y&VLT/FORS2&800&8&[9]\\
070506	&	2.308	&$	22.0	\pm	0.3 $		& $\geq-0.65 $	&	Zn&N&VLT/FORS1&800&18&[9]\\
070721B	&	3.628	&$	21.5	\pm	0.2	$	&	$\geq-2.14$	&Si&Y&VLT/FORS2&800&5&[9]\\
070802	&	2.455	&$	21.5	\pm	0.2	$	&	$\geq-0.54 $	&	Si&Y&VLT/FORS2&800&7&[9]\\
071020	&	2.145	&$	<20.30		$	&...		&	...&N&VLT/FORS2&800&5&[9]\\
071031	&	2.692	&$	22.15	\pm	0.05	$	&$-1.85 \pm0.12$&Fe&Y&VLT/UVES&55000&10&[24]\\
080210	&	2.641	&$	21.9	\pm	0.1	$	&$\geq-1.37 $&Fe&Y&VLT/FORS2&1400&25&[9][25]\\
080310	&	2.427	&$	18.7	\pm	0.1	$	&...	&...&Y&VLT/UVES&55000&30&[17]\\
080413A	&	2.433	&$	21.85	\pm	0.15 $		&	$\geq-1.56 $&Zn &N&Gemini/GMOS&1200&17&[24]\\
080603B	&	2.69	&	$ 21.85	\pm	0.05^{c}$	&	$...$	&... & Y&NOT/AlFOSC&355&...&[9]\\
080607	&	3.037	&$	22.7	\pm	0.15	$	&	$\geq-1.72$	&Fe&Y&Keck/LRIS&2000&$40$&[9][26]\\
080721	&	2.591	&$	21.6	\pm	0.1	$	&	$\geq-1.73$	&S&N&VLT/FORS1&800&40&[27]\\
080804	&	2.20542	&$	21.3	\pm	0.1	$	&		$-0.75\pm0.16$ & Zn	&N&VLT/UVES&55000&10&[9]\\
	&		& 	&		$\gtrsim-1.25$ & Zn&N&Gemini&1200&10&[10]\\
080810	&	3.35	&	$ 17.5	\pm 	0.15	$	&...	&	...&Y&NOT/AlFOSC&400&8&[28]\\
	&	 	&	 &...	&	...&Y&Keck/HIRES&50000&30&[28]\\
080913	&	6.69	&	$< 19.84$	&	...	&	...&N&VLT/FORS2&800&3&[29]\\
081008	&	1.96	&	$ 21.59	\pm	0.1	$	&	$-0.86\pm0.14 $	&S&Y&VLT/UVES&40000&5&[30]\\
 	&	 	&	 	&	$	\geq-1.41 $	&	S&Y&Gemini/GMOS&1200&12&this work\\
090205	&	4.64	&	$ 20.73	\pm	0.05	$	&	$	>-0.57$	&S&Y&VLT/FORS1&440&5&[31]\\
090323	&  3.5778	& $>19.90$&...  & ...&Y&VLT/FORS2&1200&24&[32][33]\\
090426	&	2.609	&$	19.1	\pm	0.15	$	&	...	&...&N&Keck/LRIS&1200&7&[20]\\
090516	&	4.109	&$	21.73	\pm	0.1 $		&$\geq-1.36$&Si&Y&VLT/FORS2&800&70&[34]\\
090809 	&   2.73		&$21.40\pm0.08$&	$-0.57 \pm0.10$ & Si &Y&VLT/X-Shooter&8000&12&[35]\\
090812	&	2.425	&$	22.3	\pm	0.1	$	& $\geq-1.64$&Si&Y&VLT/FORS2&800&60&[36]\\
090926A	&	2.1062	&$	21.73	\pm	0.07 $		&	$\approx -1.9$	&	S&Y&VLT/FORS2&780&35&[37]\\
&		&$			$	&$-2.18\pm0.12$	&S&Y&VLT/X-Shooter&10000&20&[37]\\
100219A	&	4.667	&$	21.13	\pm	0.12	$	&	$-0.95\pm 0.18$	&S&Y&VLT/X-Shooter&6000&6&[38]\\
	&		&	&$\geq-1.8$	&S&Y&Gemini/GMOS&1200&7&this work\\ 
100425A	&  1.755		&$21.05\pm0.10$	&$-0.96\pm 0.42$  & Fe&Y&VLT/X-Shooter&8000&4&[35]\\
110205A	&	2.214	&$	21.45	\pm	0.2	$	&$\geq-0.82$	&S&Y&FAST&2400&30&[39]\\
111008A	&	4.98968&	$ 22.3	\pm	0.06		$&	$-1.63\pm0.13$	&Fe&Y&VLT/X-Shooter&10000&10&[40]\\
111107A	&	2.893	&$21.0\pm0.10	$	&$\geq-0.45$	&S&Y&VLT/X-Shooter&8000&5&this work\\
120327A	&	2.813	&$	22.01\pm	0.09	$	&$-1.51\pm0.11$	&S&Y&VLT/X-Shooter&8000&30&[41]\\
120716A	&	2.487	&$21.55	\pm0.15$	&$\geq-1.76$	&Fe&Y&VLT/X-Shooter&8000&7&this work\\  
120815A	&	2.3574	&	$ 21.95	\pm	0.1		$&	$-0.93\pm0.13$	&Zn&Y&VLT/XShooter&10000&12&[42]\\
120909A	&	3.9293	&$	21.20\pm0.10		$	&$-0.66\pm0.11$	&S&Y&VLT/X-Shooter&8000&9&this work\\ 
121024A	&	2.2977&$	21.50\pm0.10	$	&$-0.40\pm0.12$	&Zn&Y&VLT/X-Shooter&8000&15&[43]\\ 
121201A	&	3.385	&$	21.7	\pm	0.2	$	&$...$&...&...&VLT/X-Shooter&8000&$\lesssim3$&this work\\ 
130408A	&	3.757	&$21.70	\pm0.10	$	&$-1.24\pm0.12$	&S&Y&VLT/XShooter &8000 &$50 $&this work\\ 
 		&			& 				&$\geq-1.1$	&S&Y&Gemini/GMOS&1200&$20$&this work\\
130505A	&	2.2687  &   $	20.65	\pm0.10$	&$\geq-1.42$	&Fe&Y&Gemini/GMOS&1200&$30$&this work\\ 
130606A	&	5.9134	&$	19.93	\pm	0.2	$	&...&	...&Y&Gemini/GMOS&1200&...&[44]\\
140226A  &  1.9733  & $20.60\pm 0.20 $  & $\geq-0.54$ & Fe& N & keck/LRIS & 1200&30&this work\\ 
140311A & 4.953  & $21.80\pm0.30$  &$\geq-1.65$&Ni&Y&Gemini/GMOS &1200&7&this work\\ 
140419A & 3.961  &$19.3\pm 0.2$&$...$&...&Y&Gemini/GMOS &1200&20&this work\\ 
140423A & 3.258  & $20.45\pm0.20$  &$\geq-1.44$&Fe&Y&Gemini/GMOS &1200&20&this work\\
140518A & 4.7055  & $21.65\pm0.20$  &$\geq-1.06$&S&Y&Gemini/GMOS &1200&50&this work\\
\enddata
\tablecomments{\,
List of the GRB-DLAs identified to date along GRB lines of sight. Missing metallicities are due to
lack of metal line transitions or the low signal-to-noise of the spectra. When multiple measurements for the same line of sight are listed, the one derived from
high resolution instrument is adopted in our analysis. Some values, either of \nh\ or $[X/H]$, are adopted from the literature reference listed in the last column.
References: 
[1]\cite{Castro:2003aa};
[2]\cite{Vreeswijk:2006aa};
[3]\cite{Hjorth:2003aa};
[4]\cite{Savaglio:2002aa};
[5]\cite{Fiore:2005aa};
[6]\cite{Shin:2006aa};
[7]\cite{Vreeswijk:2004aa};
[8]\cite{Jakobsson:2004aa};
[9]\cite{Fynbo:2009aa};
[10]\cite{Laskar:2011aa};
[11] \cite{Watson:2006aa};
[12]\cite{Berger:2006aa};
[13]\cite{DElia:2007aa};
[14]\cite{Prochaska:2007ab};
[15]\cite{Kawai:2006aa};
[16]\cite{Chen:2007aa};
[17]\cite{Fox:2008aa};
[18]\cite{Penprase:2006aa};
[19]\cite{Chary:2007aa};
[20]\cite{Thone:2010aa};
[21]\cite{Ferrero:2009aa};
[22]\cite{Jakobsson:2006aa};
[23]\cite{Ruiz-Velasco:2007aa};
[24]\cite{Ledoux:2009aa};
[25]\cite{De-Cia:2011aa};
[26]\cite{Prochaska:2009aa};
[27]\cite{Starling:2009aa};
[28]\cite{Page:2009aa};
[29] \cite{Patel:2010aa};
[30]\cite{DElia:2011aa};
[31]\cite{DAvanzo:2010aa};
[32]\cite{Cenko:2011aa};
[33]\cite{Savaglio:2012ab};
[34]\cite{de-Ugarte-Postigo:2012aa};
[35]\cite{skula10};
[36]\cite{Rau:2010aa};
[37]\cite{DElia:2010aa};
[38]\cite{Thone:2013aa};
[39]\cite{Cucchiara:2011ab};
[40]\cite{Sparre:2014aa};
[41]\cite{DElia:2014aa};
[42]\cite{Kruhler:2013aa};
[43]\cite{Friis:2014aa};
[44]\cite{Chornock:2013aa}.
}
\tablenotetext{a}{Metallicities are relative to solar: $[X/H]={\rm log} [X/H]_{DLA}$-{\rm log}$[X/H]_{\odot}$. For the sub-DLA systems, with $10^{19} <\nh < 2 \NHunits$, we defer the reader to future paper.}
\tablenotetext{b}{On average over multiple spectra, see Appendix A}
\tablenotetext{c}{Value reported in the literature}
\end{deluxetable}